\documentclass[authoryear,a4paper,12pt]{article}

\usepackage{lineno,hyperref,amsmath,amsfonts,float,graphicx,natbib,authblk,booktabs,enumitem}
\usepackage{csquotes}
\usepackage{xcolor,colortbl}
\usepackage[margin=1in]{geometry}

\renewcommand{\(}{\left(}
\renewcommand{\)}{\right)}
\renewcommand{\b}{\bfseries}
\newcommand{\specialcell}[2][l]{%
  \begin{tabular}[#1]{@{}l@{}}#2\end{tabular}}

\usepackage{color,colortbl}
\definecolor{LightCyan}{rgb}{0.8,1,1}
\newcommand{\clrone}{\rowcolor{lightgray!50}}
\definecolor{LightCyan2}{rgb}{0.1,1,1}
\newcommand{\clrtwo}{\rowcolor{lightgray}}
\newcommand{\colarray}{\arrayrulecolor{lightgray}}

\title{Forecasting under model uncertainty: Non-homogeneous hidden Markov models with P\'{o}lya-Gamma data augmentation}
\date{}
\author[1,*]{Constandina Koki}
\author[2]{Loukia Meligkotsidou}
\author[1]{Ioannis Vrontos}
\affil[1]{Athens Univeristy of Economics and Bussiness, 76 Patission Str, Athens, Greece}
\affil[2]{National and Kapodistrian University of Athens, Panepistimioupolis, Athens, Greece}
\affil[*]{Corresponding author: Constandina Koki, kokiconst@aueb.gr}
\begin{document}

	\maketitle

\begin{abstract}
		We consider two-state Non-Homogeneous Hidden Markov Models (NHHMMs) for forecasting univariate time series. Given a set of predictors, the time series are modeled via predictive regressions with state dependent coefficients and time-varying transition probabilities that depend on the predictors via a logistic function. In a hidden Markov setting, inference for logistic regression coefficients becomes complicated and in some cases impossible due to convergence issues. In this paper, we aim to address this problem using a new latent variable scheme that utilizes the P\'{o}lya-Gamma class of distributions. We allow for model uncertainty regarding the predictors that affect the series both linearly -- in the mean -- and non-linearly -- in the transition matrix. Predictor selection and inference on the model parameters are based on a MCMC scheme with reversible jump steps. Single-step and multiple-steps-ahead predictions are obtained by the most probable model, median probability model or a Bayesian Model Averaging approach. Using simulation experiments, we illustrate the performance of our algorithm in various setups, in terms of mixing properties, model selection and predictive ability. An empirical study on realized volatility data shows that our methodology gives improved forecasts compared to benchmark models.
\end{abstract}


\section{Introduction}\label{sec1}
	Discrete-time finite state-space Homogeneous Hidden Markov Models (HHMMs) have been extensively studied and used to model stochastic processes that consist of an observed process and a latent (hidden) sequence of states which is assumed to affect the observation sequence, see for example \cite{Ca05} and \cite{Bi99}. 
	Bayesian inference, using Markov Chain Monte Carlo (MCMC) techniques, has enhanced the applicability of HHMMs and has led to the construction of more complex model specifications including Non-Homogeneous Hidden Markov Models (NHHMMs). Initially, \cite{Di94} studied the two state Gaussian NHHMMs where the time varying transition probabilities were modeled via logistic functions. Their approach was based on the Expectation-Maximization algorithm (EM). \cite{Fi98} adopted a Bayesian perspective to overcome technical and calculation issues of classical approaches. Since then, various Bayesian methods have been proposed in the literature. For example, \cite{Sp06} modeled the time-varying transition probabilities via a logistic function depending on exogenous variables and performed model selection based on the Bayes factor. In the same spirit, \cite{Me11} considered an $m$-state NHHMM and assumed that the elements of the transition matrix are linked through exogenous variables with a multinomial logistic link, whereas the observed process conditional on the unobserved process follows an autoregressive model of order $p$. They accommodated and exploited model uncertainty within their Bayesian model -- by allowing covariate selection only on the transition matrix -- to improve the predictive ability of NHHMMs on economic data series.
	 \par
	 Based on experimental evidence, the algorithm of \cite{Me11} (M\&D) faces convergence issues when there exists model uncertainty, due to the data augmentation scheme of \cite{Ho06}. \cite{Po13} confirm the efficiency issues in the \cite{Ho06} scheme and propose a P\'{o}lya-Gamma data augmentation strategy that significantly improves over various benchmarks, e.g., \cite{OBr04,Fr10,Fu13}. Furthermore, the recent work of \cite{Ho17} confirms that using P\'{o}lya-Gamma data augmentation to parametrize the transition probabilities of NHHMMs results in an algorithm that mixes well and provides adequate estimates of the model parameters. 
	 \par
	 Motivated by this, we revisit the work of \cite{Me11} by employing the recent methodological advances on the P\'{o}lya-Gamma data augmentation scheme of \cite{Po13}. We consider two-state NHHMMs (easily extended to m-state NHHMMs) in which the time series are modeled via different predictive regression models for each state, whereas the transition probabilities are modeled via logistic regressions. Given an available set of predictors, we allow for model uncertainty regarding the predictors that affect the series both linearly -- directly in the mean regressions -- and non-linearly -- in the transition probability matrix. 
	  \par
	 The resulting model is a Non-Homogeneous Polya-Gamma Hidden Markov Model, which we will denote by NHPG. Bayesian inference is performed via a MCMC scheme which overcomes difficulties and convergence issues inherent in existing MCMC algorithms. To this end, we exploit the missing data representation of hidden Markov models and construct an MCMC algorithm based on data augmentation, consisting of several steps. First, we sample the latent sequence of states via the Scaled Forward-Backward algorithm of \cite{Sc02}, which is a modification of the Forward-Backward algorithm of \cite{Ba70} who used it to implement the classical EM algorithm. Then, we use a logistic regression representation of the transition probabilities and simulate the parameters of the mean predictive regression model for each state, via Gibbs sampling steps. Finally, we incorporate variable selection within our MCMC scheme by using the Reversible Jump (RJ) algorithm model of \cite{Gr95,Ha12}.
 
	\par Different approaches have been used in the literature to cope with the model selection problem.  The use of information criteria, such as Akaike's Information Criterion (AIC, \cite{Ak73}), the Bayesian Information Criterion (BIC) of \cite{Sc78}, the Deviance Information Criterion  (DIC, \cite{Sp02}) or the Widely applicable Bayesian Information Criterion (WBIC, \cite{Wa13}), is another approach to variable selection. A study for comparing variable selection methods is well presented in \cite{O'Hara09} whilst \cite{De02} study the variable selection methods in the context of model choice. \cite{Ho17} consider a NHHMM similar to ours for modeling multivariate meteorological time series data. In that paper, the transition probabilities are modeled via multinomial logistic regressions affected by a specific set of exogenous variables. The authors use the BIC criterion for choosing the best model among a pre-specified class of models. We extend this work by considering the problems of statistical inference and variable selection jointly, in a purely Bayesian setting. The proposed model is flexible, since we do not decide a priori which covariates affect the observed or the unobserved process. Instead, we have a common pool of covariates $\left\{X\right\}$ and within the MCMC algorithm, we gauge which covariates are included in subset $\left\{X^{(1)}\right\}$ affecting the mean predictive equation of the observed process, and which covariates are included in subset $\left\{X^{(2)}\right\}$ affecting the time-varying transition probabilities.
	\par Our probabilistic approach is based on the calculation of the posterior distribution of different NHPGs. Posterior probabilities can be used either for selecting the most probable model (i.e., making inference using the model with the highest posterior probability), or for Bayesian model averaging (i.e., producing inferences averaged over different NHPGs). \cite{Ba04} argue that the optimal predictive model is not necessarily the model with the highest posterior probability but the median probability model, which is defined as the model consisting of those covariates which have overall posterior probability of being included in the model -- inclusion probability -- greater or equal to 0.5. We calculate both the posterior probabilities of the models and the probabilities of inclusion. \par 
    We use our model for predicting realized volatility. Accurate forecasting of future volatility is important for asset allocation, portfolio construction and risk management, see \cite{Go06}. A review on the realized volatility literature can be found in \cite{Mc08}. The relationship between the volatility and macroeconomic and/or financial variables is investigated in \cite{Pa12,Ch12,Me19} among others. The proposed NHPG captures not only the linear relationship between the logarithm of realized volatility and a set of predictors, as in the model of \cite{Ch12} (CSS), but also the nonlinear relationship, as well as other special characteristics of the analyzed series, such as heteroscedasticity and autocorrelation. NHPG outperforms the M\&D, CSS models and the HHMM, in terms of forecasting ability. 
	\par The MCMC output of the predictive density of the NHPG is multimodal and thus, scoring rules that are not sensitive to distance should be avoided (\cite{Gn07}). For instance the logarithmic scoring rule gives harsh penalty for low probability events (\cite{Bo11, Gn07}) and prefers the forecast density that is less informative (\cite{Ma13}). In this case, a better alternative not only for validating the model performance but also for assessing the quality of forecasts is the Continuous Rank Probability Score (CRPS). This proper scoring rule has gained a lot of interest in the meteorological community, see \cite{Gr06}, and proves to be the most appropriate rule also for the NHPG model. 
	 \par
	In summary, the main contributions of our paper are the following
	\begin{enumerate}[itemsep=0pt]
		\item We propose a flexible model (NHPG) that can detect the linear and a non-linear relationship between the predictors and the studied time series. This results in a stable algorithm which does not need tuning and can be used as a black box for predicting time series.
		\item  We present experimental evidence in support of the claim that the NHPG model has an improved performance in terms of  variable selection and forecasting ability when compared with M\&D. This is at no cost of computational complexity and running time.
		\item We provide evidence that the proposed algorithm performs well also with real datasets by obtaining improved forecasts on the realized volatility data set of \cite{Ch12}. 
	\end{enumerate}
	\par The paper proceeds as follows: In Section~\ref{The model}, we outline the proposed model and in Section~\ref{Inference}, we describe our Bayesian computational strategy both with and without model uncertainty. Section~\ref{Forecasting Criteria}, presents our forecasting criteria. Section~\ref{Simulation Study} contains numerical experiments and Section~\ref{Empirical Application} proceeds with the main application on the realized volatility data set. Finally, Section~\ref{Conclusions} concludes the paper. A case study with without uncertainty as well as further details on the metrics of comparisons and benchmark models are deferred to the Appendix.
\section{The Non-Homogeneous P\'{o}lya-Gamma Hidden Markov Model}\label{The model}
The proposed Non-Homogeneous P\'{o}lya-Gamma hidden Markov model (NHPG) for univariate time series is described as follows. Consider an observed random process $\left\{Y_{t}\right\}$ and a hidden underlying process $\left\{Z_{t}\right\}$ which is a two-state non-homogeneous discrete-time Markov chain that determines the states of the observed process. Let $y_{t}$ and $z_{t}$ be the realizations of the random processes $\left\{Y_{t}\right\}$ and $\{Z_t\}$, respectively. We assume that at time $t,\ t=1,\dots,T$, $y_{t}$ depends on the current state $z_{t}$ and not on the previous states. Consider also a set of $r-1$ available predictors $\left\{X_{t}\right\}$ with realization $x_{t}=(1,x_{1t},\dots ,x_{r-1t})$ at time $t$. A subset of the predictors $X_{t}^{(1)}\subseteq \left\{X_{t}\right\}$ of length $r_1-1$ is used in the regression model for the observed process and a subset $X_{t}^{(2)}\subseteq \{X_{t}\}$ of length $r_2-1$ is used to describe the dynamics of the time-varying transition probabilities. Thus, we allow the covariates to affect the observed process $\{Y_{t}\}$ non-linearly. 
\par The observed random process $\left\{Y_{t}\right\}$ can be written in the form $$Y_t=g(Z_{t})+\epsilon_t,$$ 
where $g(Z_t)=X^{(1)}_{t-1}B_{Z_t}$ is a linear function, $B_{Z_t}=(b_{0Z_t},b_{1 Z_t},\dots ,b_{r_1-1Z_t})'$ are the regression coefficients and $\epsilon_t\sim \mathcal{N}(0,\sigma^2_{Z_t})$. We use $\mathcal{N}(\mu,\sigma^2)$ to denote the normal distribution with mean $\mu$ and variance $\sigma^2$. In a less formal way, if $s$ represents the hidden states, the observed series given the unobserved process has the form  
$$Y_t\mid Z_t=s \sim \mathcal{N}(X^{(1)}_{t-1}B_s,\sigma^2_{s}), \;s=1,2.$$
The dynamics of the unobserved process $\left\{Z_{t}\right\}$ can be described by the time-varying transition probabilities, which depend on the predictors $X_{t}^{(2)}$ and are given by the following relationship 
	$$P(Z_{t+1}=j\mid Z_t=i)=p^{(t)}_{ij}=\frac{\exp(x^{(2)}_{t}\beta_{ij})}{\sum^{2}_{j=1}\exp(x^{(2)}_t\beta_{ij})}, \; i,j=1,2,$$ 
where $\beta _{ij}=(\beta _{0,ij},\beta _{1,ij},\dots ,\beta_{r_2-1,ij})^{\prime }$ is the vector of the logistic regression coefficients to be estimated. Note that for identifiability reasons, we adopt the convention of setting, for each row
of the transition matrix, one of the $\beta_{ij}$ to be a vector of zeros. Without loss of generality, we set $\beta_{ij}=\beta_{ji}=\mathbf{0}$ for $i,j=1,2, i\neq j$. Hence, for $\beta_i=\beta_{ii},\;i=1,2$ probabilities can be written in a simpler form $$p^{(t)}_{ii}=\frac{\exp(x^{(2)}_{t}\beta_{i})}{1+\exp(x^{(2)}_t\beta_{i})} \ \text{and}\   p^{(t)}_{ij}=1-p^{(t)}_{ii} ,\ i,j=1,2,\ i\neq j.$$ 
The unknown quantities of the NHPG are $\left\{\theta_s=\(B_{s},\sigma _{s}^{2}\),\beta_s, s=1,2 \right\}$, i.e., the parameters in the mean predictive regression equation and the parameters in the logistic regression equation for the transition probabilities of the unobserved process $\left\{Z_{t}\right\}$, $t=1,...,T$. Our model and the methods developed in this paper can be easily generalized into an m-state NHHMM, where the rows of the transition matrix are modeled by multinomial logistic regressions.
\section{Bayesian Inference and Computational Strategy}\label{Inference}
The key steps in our proposed framework are the following. First, for a given NHPG, we construct a Markov chain which has as stationary distribution the posterior distribution of the model parameters. Simulation of this Markov chain provides, after some burn in period and adequately many iterations, samples from the posterior distribution of interest; see, for details, \cite{Be95}. Second, for a given set of competing models, each including a different set of predictors in the mean regression and/or in the transition probabilities equation, we base our inference about the models on their posterior probabilities. This improves over the approach which considers the models separately and chooses the best model via significance tests or via model selection criteria.
\subsection{The MCMC Sampling Scheme}
The main steps of the proposed MCMC algorithm for joint inference on model specification and model parameters are the following.
\begin{enumerate}[noitemsep]
	\item Start with initial values of $\beta, \theta=\(B,\sigma^2\)$.
	\item Calculate the probabilities of the time-varying transition matrix.
	\item Given the model's parameters, simulate the hidden states  using a Scaled Forward-Backward (\cite{Sc02}) algorithm.
	\item Simulate the mean regression parameters via a Gibbs sampler method.	
	\item Simulate the coefficients $\beta$  using the P\'{o}lya-Gamma representation by \cite{Po13}.
	\item Use a double reversible jump algorithm to update the set of covariates that affect the transition matrix and those that affect the mean regression model.
	\item Make one-step-ahead predictions conditional on the simulated unknown quantities.
	\item Repeat steps 3-6 until convergence and then repeat steps 3-7.
\end{enumerate} 
In the next subsections, we present each step in detail.

\subsection{Inference for fixed sets of predictors}\label{Inference fixed}
For a given NHPG, i.e., for fixed sets of predictors used in the mean equation and the transition probabilities $X^{(1)}$ and $X^{(2)}$, respectively, we update in turn (i) the latent variables $z^{T}$ given the current values of the model parameters by using the scaled Forward-Backward algorithm (\cite{Sc02}) (ii) the logistic regression coefficients by adopting the auxiliary variables method of \cite{Po13} given the sequence of states $z^T$, and (iii) the mean regression coefficients conditional on $z^T$ by using the Gibbs sampling algorithm. 
\par Let $y^{T}=(y_{1},\dots ,y_{T})$ be the history of the observed process, $z^{T}=(z_{1},\dots ,z_{T})$ the sequence of states up to time $T$, and let $f_{s}(\cdot)$ denote the normal probability density function of $Y_{t}\mid Z_{t}=s$, $s=1,2$ and $\pi_{1}(z_{1})$ the initial distribution of $Z_1$. The joint likelihood function of the data, $y^{T}$, and the sequence of states, $z^{T}$, is given by 
\begin{align*}
\mathcal{L}\(\theta,\beta\)&= \pi(y^{T},z^{T}\mid X,\theta ,\beta )=\pi (y^{T}\mid \ z^{T},X,\theta, \beta )\pi (z^{T}\mid X,\theta, \beta )\\
	&=\pi	_{1}(z_{1})f_{z_{1}}(y_{1})\prod_{t=2}^{T}p_{z_{t-1}z_t}^{(t-1)}f_{z_t}(y_{t})\\
	&=\prod_{i=1}^{2}\prod_{j=1}^{2}\left[\prod_{t:z_{t=j}}p_{ij}^{(t-1)}\right]
	\( \frac{1}{2\pi \sigma _{j}^{2}}\) ^{N_{j}/2}\exp
	\left\{ -\frac{1}{2\sigma _{j}^{2}}(Y_{j}-X_{j}^{(1)^{\prime
	}}B_{j})^{\prime }(Y_{j}-X_{j}^{(1)^{\prime }}B_{j})\right\}.
\end{align*}
We use the notation $N_s,\; s=1,2$ for the number of times the chain was in state $s$, that is $N_{s}=\sum_{t=1}^{T}I(Z_{t}=s)$, with $I$ the indicator function. If a prior distribution $\pi\(\theta,\beta\)$ is specified for the model parameters, then inference on all the unknown quantities in the model is based on their joint posterior distribution $\pi (\theta ,\beta ,z^{T}\mid y^{T})\propto \pi (\theta ,\beta )\pi(y^{T},z^{T}\mid \theta ,\beta )$.\par
For the parameters in the mean predictive regression equation, we use conjugate prior distributions, i.e., $ \sigma _{s}^{2}\sim \mathcal{IG}(p,q)\text{, }B_{s}\mid\sigma _{s}^{2}\sim 
\mathcal{N}(L_0 ,\sigma _{s}^{2}V_0 ) ,\ s=1,2,$ where $\mathcal{IG}$ denotes the Inverted-Gamma distribution. After some straightforward algebra we derive the marginal posterior distribution for the state specific parameters $\sigma_s$ and conditional posterior distribution for $B_s$,  
$$\sigma _{s}^{2}\mid y^{T},z^{T}\sim \mathcal{IG}\left( p+\frac{n_{s}}{2},q+\frac{1}{2}\(L_{0s}^{\prime }V^{-1}_{0s} L_{0s} +Y_{s}^{\prime}Y_{s}-L^{\prime}_{s}V_{s}^{-1}L_{s}\)\right),$$
$$B_{s}\mid\sigma _{s}^{2},z^{T},y^{T}\sim \mathcal{N}\(L_{s},\sigma _{s}^{2}V_s\)
,$$
with $V_{s}=\left( V_{0s}^{-1}+X_{s}^{(1)^{\prime }}X_{s}^{(1)}\right) ^{-1} \text{and} \ L_{s}=V_{s}\left( V_{0s} ^{-1}L_{0s} +X_{s}^{(1)^{\prime }}Y_{s}\right).$
\par
 To make inference about the logistic regression coefficients, we use the auxiliary variables method of \cite{Po13} as described in Subsection~\ref{Polya-Gamma}. Given the auxiliary variables $\omega_s$, a conjugate prior for the logistic regression coefficients $\beta _{s}$, $s=1,2$ is multivariate normal distribution $\mathcal{N}\(m_{\beta_s} ,V_{\beta_s}\)$. The conditional posterior distribution of $\beta _{s}$, $s=1,2$ is again a multivariate normal, see Section~\ref{Polya-Gamma}. \par
\subsubsection{Simulation of the logistic regression coefficients}\label{Polya-Gamma}
We model the two diagonal elements of probability transition matrix by linking them to the set of covariates using a logistic link. We use the data augmentation scheme of \cite{Po13} since, as shown in their work, the estimation of logistic regression coefficients using this scheme is superior, in terms of efficiency.
\par Given the unobserved (latent) data $z^{T}=(z_{1},\dots ,z_{T})$ we define, for $t=1\dots,T-1$, the quantity $\tilde{Z}_{t+1}^{s}=I\left[ Z_{t+1}=Z_{t}=s\right]$. The sum $\sum_t\tilde{Z}_{t+1}^{s}$, is the number of times that the chain was at the same state for two consecutive time periods. Then,  $$p\(\tilde{Z}_{t+1}^{s}=1\mid x^{(2)}_{t}\)=p^{t}_{ss}=\frac{\exp\(x^{(2)}_{t}\beta_{s}\)}{1+\exp\(x^{(2)}_{t}\beta_{s}\)} \Leftrightarrow logit(p^{t}_{ss})=x^{(2)}_{t}\beta_{s}, \ s=1,2.$$
\cite{Po13} proved that binomial likelihoods -- thus Bernoulli likelihoods in our simpler case -- parametrized by log odds can be represented as mixtures of Gaussian distributions with respect to the P\'{o}lya-Gamma distribution. The main result of \cite{Po13} is that letting $p(\omega)$ be the density of a latent variable $\omega$ with $\omega\sim \mathcal{PG}(b,0)$, for $b>0$, the following integral identity holds for all $a\in \mathbb{R}$
$$\frac{\exp\(\psi\)^{a}}{\(1+\exp\(\psi\)\)^b}=2^{-b}\exp\(k\psi\)\int_{0}^{\infty} \exp\(-\omega\psi^2/2\)p\(\omega\)d\omega, $$
where $k=a-b/2$. Furthermore, the conditional distribution of $\omega\mid \psi$ is also P\'{o}lya-Gamma, $\mathcal{PG}(b,\psi)$. Using the previous result and setting $\Omega_s=diag\{ \omega_{1,s},\dots,\omega_{N_{s},s} \}$ as a set of latent variables, the likelihood for each state $s=1,2$ is 
\begin{align*}
\mathcal{L}\(\beta_s,\omega_s\)&=\prod_{t=1}^{N_s}\left\{\frac{\exp\(x^{(2)}_{t}\beta_{s}\)}{1+\exp\(x^{(2)}_{t}\beta_{s}\)}\right\}^{\tilde{z}_t}\left\{\frac{1}{1+\exp\(x^{(2)}_{t}\beta_{s}\)}\right\}^{1-\tilde{z}_t}\\
&\propto \prod_{t=1}^{N_s} \exp\(k_t x^{(2)}_{t}\beta_{s}\)\int_{0}^{\infty}\exp\left\{-\omega_{t,s} \(x^{(2)}_{t}\beta_{s}\)^2/2\right\}p(\omega_{t,s})d\omega_{t,s}.
\end{align*}
Conditioning on $\Omega_s$, one can derive the expression 
$$\pi\(\beta\mid z^t,\omega_s\)\propto \pi\(\beta\)\prod_{t=1}^{N_s}\exp\left\{-\frac{\omega_{t,s}}{2}\(\(x^{(2)}_{t}\beta_{s}\)^2-\frac{2k_tx^{(2)}_{t}\beta_{s}}{\omega_{t,s}}+\frac{k^{2}_{t}}{\omega^{2}_{t,s}}\)\right\}.$$
Assuming as prior distributions $\omega \sim \mathcal{PG}(b,0)$ and $\beta\sim \mathcal{N}\(m_{\beta_0},V_{\beta_0}\)$, simulation from the posterior distribution can be done iteratively in two steps:
\begin{gather*}
\omega_{t,s}\mid\tilde{z_t}\sim \mathcal{PG}\(1,x^{(2)}_{t}\beta_{s}\),\  t=1:N_s,\ s=1,2,\\
\beta_{s}\mid\tilde{Z},\Omega_s \sim \mathcal{N}(m_{\omega_s},V_{\omega_s}),\\
V_{\omega_s}=\left( X^{(2)\prime} \Omega_s X^{(2)} +V^{-1}_{\beta_0} \right)^{-1} \text{and}\  m_{\omega_s}=V_{\omega_s}\left(X^{(2)\prime}k+V^{-1}_{\beta_0}m_{\beta_0}\right),
\end{gather*}
where $\mathcal{PG}$ denotes the P\'{o}lya-Gamma distribution and $k=\left( \tilde{z}_1-1/2,\dots, \tilde{z}_{N_s}-1/2\right)$.
\subsection{Inference under model uncertainty}
We consider the full model comparison problem. The uncertainty about which predictors should be included in the mean regression model and in the transition probability equation is treated using a double RJMCM algorithm. 
In this setting, the RJMCMC does not need tuning and hence it can be used as a black box.

\par
\par Suppose that a prior $\pi\(k\)$ is specified over k models $\(M_1,M_2,\dots, M_k\)$ in a countable set $\mathcal{K}$ and for each $k$ we are given a prior distribution $\pi\(\theta_k \mid k\)$ along with a likelihood $\mathcal{L}\(y\mid \theta_k,k\)$ for data y. The joint prior for $\theta_k$ and $k$ is $\pi(k,\theta_k) =\pi(\theta_k\mid k)\pi\(k\)$. When a move of type $m$ from $\tilde{x}=\(k,\theta_k\)$ to $\tilde{x}^{\ast}=\(k^{\ast},\theta^{\ast}_{k^{\ast}}\)$ is proposed from the proposal distribution $g$ and if $j_m(\tilde{x})$ denotes the probability that move $m$ is attempted at state $\tilde{x}$ and $j_{m^{\ast}}(\tilde{x}^{\ast})$ the probability of the reverse move, we accept the proposed move with probability $\alpha_m\(\tilde{x},\tilde{x}^{\ast}\)=\min{\left\{1,A_m(\tilde{x},\tilde{x}^{\ast})\right\}}$ where
$$A_m(\tilde{x},\tilde{x}^{\ast})=\frac{\mathcal{L}\(y^T\mid\tilde{x}^{\ast}\)\pi\(\theta^{\ast}_{k^{\ast}}\mid k^{\ast}\)\pi\(k^{\ast}\)j_{m^{\ast}}(\tilde{x}^{\ast})g'_m\(u^{\ast}\mid \tilde{x}^{\ast},k\)}{\mathcal{L}\(y^T\mid \tilde{x}\)\pi\(\theta_k\mid k\)\pi\(k\)j_m\(\tilde{x}\)g_m\(u\mid \tilde{x},k^{\ast}\)}\left|\frac{\partial\(\theta^{\ast}_{k^{\ast}},u^{\ast}\)}{\partial\(\theta_{k},u\)} \right|,$$
and $\left|\frac{\partial\(\theta^{\ast}_{k^{\ast}},u^{\ast}\)}{\partial\(\theta_{k},u\)} \right|$ is the Jacobian of the transformation.
\par
In each step, we choose to add or remove one covariate with probability $0.5$. Then, we randomly choose which covariate will be added or removed from the corresponding set of the non-included or included covariates. We propose a new value for the mean equation coefficients $B^{\ast}$ or for the regression equation coefficients $\beta^{\ast}$ from the full conditional posterior density, conditionally on the other coefficients. Thus, the Jacobian of the transformation will be equal to unity. To be more specific, if we want to update the covariates in the mean equation, the proposal distribution $g'$ is just the product of the two conditional posterior distributions. With some straightforward matrix algebra, the acceptance probability for the mean equation is $\alpha_B=\min{\left\{1,A_B\right\}}$ and the acceptance probability for the transition matrix is $\alpha_{\beta}=\min{\left\{1,A_{\beta}\right\}}$ where   
\begin{align*}
A_B=&\frac{j_{m^{\ast}}\(k^{\ast}\)}{j_m\(k\)}\prod^{2}_{s=1}\frac{ \left|V^{\ast}_s\right|^{1/2}\left|V_{0s}\right|^{1/2}}{\left|V_s\right|^{1/2}\left|V^{\ast}_{0s}\right|^{1/2}}\\
&\times\exp \left\{ -\frac{1}{2\sigma^{2}_{s}} \(L^{\ast'}_{0s}V^{\ast-1}_{0s}L^{\ast}_{0s}-L^{\ast'}_sV^{\ast-1}_sL^{\ast}_s-L^{\prime}_{0s}V^{-1}_{0s}L_{0s}+L^{\prime}_sV^{-1}_sL_s\)\right\}
\end{align*}
and
\begin{align*}
A_{\beta}=&\frac{j_{m^{\ast}}\(k^{\ast}\)}{j_m\(k\)}\prod^{2}_{s=1}\frac{ \left|V^{\ast}_{\omega s}\right|^{1/2}\left|V_{\beta_0s}\right|^{1/2}}{\left|V_{\omega s}\right|^{1/2}\left|V^{\ast}_{\beta_0s}\right|^{1/2}}\\
&\times \exp \left\{ -\frac{1}{2\sigma^{2}_{s}} \(L^{\ast'}_{\beta_0s}V^{\ast-1}_{\beta_0s}L^{\ast}_{\beta_{0}s}-L^{\ast'}_\omega V^{\ast-1}_{\omega s}L^{\ast}_{\omega s}-L^{\prime}_{\beta_0s}V^{-1}_{\beta_0s}L_{\beta_0s}+L^{\prime}_{\omega s}V^{-1}_{\omega s}L_{\omega s}\)\right\}.
\end{align*}\par Having described the inference and model selection of our model, we can now proceed to the description of the forecasting methodology.
\section{Bayesian Forecasting and Scoring rules}\label{Forecasting Criteria}
The posterior predictive density cannot be found in closed form, but can be evaluated numerically. Given model $M$, the predictive distribution of $y_{T+1}$ is
$$f_p\(y_{T+1}\mid y^T\)=\int f\(y_{T+1}\mid y^T,z^T,M,\beta_M,\theta_M\)\pi\(\beta_M,\theta_M\mid y^{T}\)d\beta_M d\theta_M,$$
where $f\(y_{T+1}\mid y^T,z^T,\beta_M,\theta_M\)=\sum_{s=1}^{2}P\(Z_{T+1}=s\mid Z_T=z_T\)f_s\(y_{T+1}\).$
In practice, we follow an iterative procedure within our MCMC algorithm to draw a sample from the posterior predictive distribution. At the $r$-th iteration of our algorithm, the algorithm chooses model $M_r$. Furthermore, the hidden states and the unknown parameters $\beta_{M_r},\theta_{M_r}$ are simulated as described in Subsection~\ref{Inference fixed}. To make an one-step-ahead prediction (i.e., simulate $y_{T+1}$), we first simulate the hidden state for time $T+1$ from the discrete distribution based on the transition probabilities $P\(Z^{(r)}_{T+1}=s\mid Z_T=z^{(r)}_{T}\), \; s=1,2$, and then, conditional on the hidden state, we draw a value $y^{r}_{T+1}$ from $\mathcal{N}\(X^{(1)}_{T}B_{s,M_r},\sigma^2_{s,M_r}\),\;s=1,2.$
Given $y_{T+l}$, $Z_{T+l}$ and the covariates $X_{T+l-1}$, for l=1,$\dots,L$, we may also update the transition matrix $P^{T+l}$, simulate $Z_{T+l+1}$ and finally simulate the prediction $y_{T+l+1}$ from its respective predictive distribution. In this way, in each iteration we obtain sequentially a sample of $L$ one-step-subsequent predictions.
\subsection{Forecasting criteria}
In our model, the predictive distributions are multimodal. Hence, to evaluate the quality of the obtained forecasts or to compare with benchmark models, the selection of the right scoring rule is integral, \cite{Ge14}. In the same manner, \cite{Ge06} observe that the predictive accuracy is valued not only for its own sake, be it can used as a metric to evaluate the model's performance.
\par Advances in numerical integration via MCMC algorithms made probabilistic forecasts possible. Besides, having the posterior predictive distribution, one can obtain point forecasts using suitable scoring functions (\cite{Gn14}). Scoring rules provide summary measures for the evaluation of probabilistic forecasts by assigning a numerical score based on the forecast and on the event or value that it materializes. We refer to \cite{Gn07, Ma13} for a review on the theory and properties of scoring rules. A widely used, extensively studied and quite powerful criterion is the Logarithmic Score (LS), see \cite{Ge14, Gs07} and references therein. It is based on the logarithm of the posterior predictive density evaluated at the observed value. However, LS lacks robustness as it involves harsh penalty for low probability events and thus is sensitive to extreme cases (\cite{Bo11}). Besides, comparing the entropies of the forecasts, \cite{Ma13} showed that LS prefers the forecast density that is less informative. In the same spirit \cite{Gn07} noticed that measures which are not sensitive to distance give no credit for assigning high probabilities to values near but not identical to the one materializing. Sensitivity to distance seems desirable when predictive distributions tend to be multimodal, which is the case of our model. To deal with this, one could calculate the Continuous Ranked Probability Score (CRPS) which is based on the cumulative predictive distribution, see Appendix~\ref{ApCRPS} for the definition. \cite{Bo11} argued that when density forecasts are collected in histogram format, then the ranked probability score has advantages over the other studied scoring rules.  
\par
To compute the CRPS for the forecast $y_l$ we use the identity of \cite{Sz05},
$$CRPS(F_p,y_l)=\frac{1}{2}E_F\left|Y-Y'\right|-E_F\left|Y-y_l\right|,$$
were $Y,Y'$ are independent copies of a random variable with the posterior predictive distribution function $F_p$ (see also \cite{Gs07}).
\par
Finally, along with the CRPS, we also use two standard point forecasting criteria: the Mean Square Forecast Error, $MSFE=\frac{1}{L}\sum^{T+L}_{l=T+1}\(y_{i}-\hat{y}_{l}\)^2$ and the Mean Absolute Forecast Error, $MAFE=\frac{1}{L}\sum^{T+L}_{l=T+1}\left| y_{l}-\hat{y}_{l}\right|$. 
The values for CRPS, MSFE and MAFE are computed in every iteration of the MCMC algorithm. In the end, we keep as CRPS, MSFE and MAFE the average over all MCMC iterations.
\section{Simulation Study}\label{Simulation Study}
We have conducted a series of simulation experiments to assess the performance of the proposed approach in terms of inference, model selection and predictive ability. We have scrutinized our algorithms, using different sample sizes and assigning various values to the parameters. Our experiments have been carried out using MATLAB 2017b on a Windows 10 system with 32GB of RAM and Intel Core i7 8-core processor.
\par
To assess its inferential ability we benchmarked our model with the N\&D model (without model uncertainty) and with a Homogeneous Hidden Markov Model (HHMM), see Appendix~\ref{ApCaseStudy}. The NHPG is at least as good as the M\&D model -- in forecasting ability, and sample quality -- but is faster and more efficient (as reported in Table~\ref{Summary_fixed}, Effective Sample Rate). 
In Table~\ref{Summary_fixed}, we present a summary of the case study of the fixed model.
\par Our model shines when there is model uncertainty, Section~\ref{Variable selection}. We compare NHPG with existing variable selection schemes, i.e., the M\&D, a HHMM with RJ step and a model using the spike and slab prior for variable selection as studied in \cite{Na14} and referred to as BAeyesian Shrinking And Diffusing priors (BASAD), see Appendix~\ref{ApBench}.
\par The data were generated either from a HHMM or from a NHHMM with covariates simulated from independent normal distributions.We found that the mean equation coefficients converged rapidly, whereas the logistic regression coefficients converged only after some burn in period. The hidden chain $Z^T$ was well estimated. For each iteration, we kept a replication of the hidden chain and compared it with the real simulated hidden chain, using a 1-0 loss function (see	Figure~\ref{model_experiment_series}).
\par Furthermore, to test the predictive ability of our model, we kept $L$ out-of-sample observations. We calculated, for all the competing models, the CRPS, the MSFE and the MAFE. However, we note that due to the large out-of-sample period, we only report the averages (for all the draws) of the aforementioned forecasting criteria. In all the experiments, we found that our model outperforms all competing models in forecasting the observed process. 
\subsection{Case study: The NHPG with model uncertainty}\label{Variable selection}
The main applications in which our algorithm considerably improves over the benchmark models -- M\&D, BASAD -- is when there exists model uncertainty. We simulated data from a NHHMM of size $T=1200$. From a common pool of independently normally distributed covariates $X=\{1,X_1,X_2,X_3,X_4,X_5,X_6,X_7,X_8,X_9\}$ with means $\mu_x=\left[4,3,-2,-5,2.5,-4,-6,7,1\right]$ and variances $\sigma^2_x=\left[1,1,0.5,1,1,1,0.5,2,1.5\right]$, we used 3 covariates $X^{(1)}=\{1,X_1,X_2,X_3\}$ affecting the mean equation and $X^{(2)}=\{1,X_1,X_2,X_4\}$ the transition matrix. The mean equation parameters were $B_1=\left[2, -0.3,2, 2\right]'$, $\sigma_1^2=1.5$ and $B_2=\left[1, 3, 4, 3\right]'$, $\sigma_2^2=0.8$ whereas the logistic regression coefficients where $\beta_1=\left[1.5,1,2, 3\right]'$ and $\beta_2=\left[3, -2.5, 4, 1\right]'$, for the two states respectively. 
\par Our results are based on a sample of 15000 predictions after discarding an initial burn in period of 10000 iterations. We kept $L=96$ out-of-sample observations and we computed a sequence of one-step-ahead forecasts of the real observed process. In this forecasting analysis, we also included the HHMM with variable selection, in the mean equation. We used non-informative priors for the unknown parameters $\sigma^2_s, B_s,\beta_{s},\ s=1,2$, that is $\sigma _{s}^{2}\sim \mathcal{IG}(0.1,0.1)$, $B_{s}\mid \sigma _{s}^{2}\sim \mathcal{N}\(0,100\sigma^2_s\times I\)$ and finally $\beta_{s} \sim \mathcal{N}\(0,100\times I\)$. Also, as suggested by \cite{Na14} and \cite{Na18}, we used as hyperparameters values $\tau^{2}_{0B,n}=\frac{\hat{\sigma}^2}{10T}, \tau^{2}_{1B,n}=\hat{\sigma}^2\text{max}\(\frac{r^{2.1}_{1}}{100T},\text{log}\(T\)\)$ and $\tau^{2}_{0\beta,n}=\frac{1}{T}, \tau^{2}_{1\beta,n}=\text{max}\(\frac{r^{2.1}_{2}}{100T},1\)$, where $\hat{\sigma}^2$ is the estimated variance of the data $Y$. 
\par
Our approach was able to identify -- as the most probable or the median probability model -- the correct data generating process. This was in contrast to the competing methodologies, as can be seen in Table~\ref{Variable_simulated}. Results from further simulation studies (not reported here), imply that the performance of our method to identify the true data generating process remains robust in the choice of  parameters. In terms of comparison, the competing algorithms could perform at most equally well.
\begin{table}[H]
	\centering
	\setlength{\tabcolsep}{10pt}
	\renewcommand{\arraystretch}{1.4}
	\begin{tabular}{lcc@{}ccc}
	\clrtwo
		&&&\multicolumn{3}{c}{\b Median probability model}\\
		&True Model&&NHPG & M\&D&BASAD \\
		\clrone
		\b ME&$X_1, X_2,X_3$&&$X_1, X_2,X_3$&$X_1, X_2,X_3$&$X_2,X_4,X_9$\\
		\b TM& $X_1, X_2,X_4$&&$X_1, X_2,X_4$
		&$X_1, X_2, \dots, X_7,X_9$&$X_1,X_2,X_4$\\	   
		 \noalign{\global\arrayrulewidth=0.5mm}
		\colarray\hline
	\end{tabular}
	\caption{Median probability models using the proposed methodology (NHPG), the methodology proposed by \cite{Me11} (M\&D) and the model of \cite{Na14} (BASAD), respectively. The first row (ME) shows the covariates used in the Mean Equation and the second row (TM) the covariates of the Transition Matrix. The proposed methodology is the only to identify the true data generating process. }
	\label{Variable_simulated}
\end{table}
\par
\begin{figure}[H]
	\centering\includegraphics[width=\linewidth, trim=2cm 4cm 2cm 1cm]{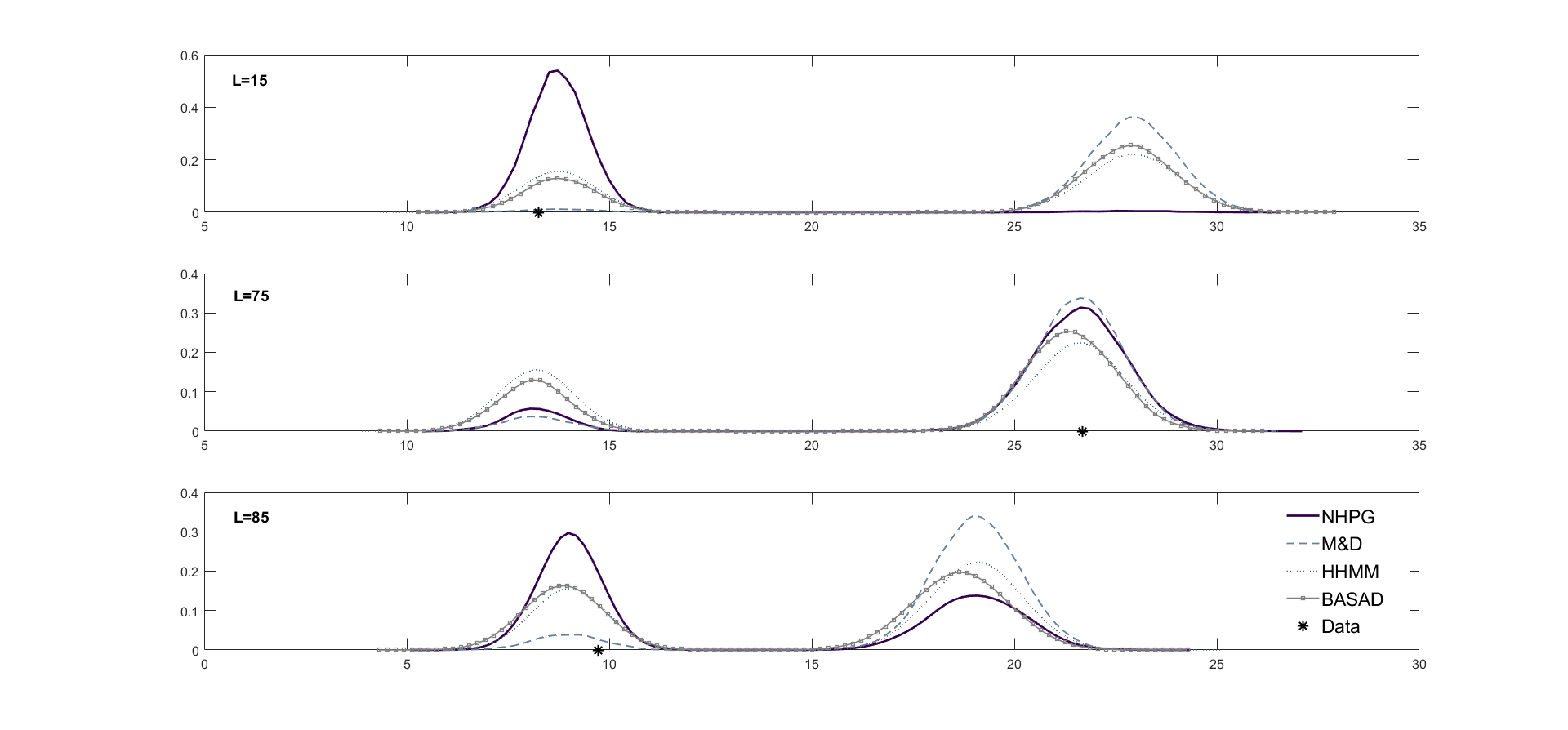}
	\caption{Plots of the empirical posterior predictive distributions based on a normal kernel function for three randomly selected out-of-sample forecasts, $L=15, 75, 85$, using the NHPG (black continuous line), M\&D (gray dashed line), the HHMM (gray dotted line) and the BASAD (gray squared line). Actual out-of-sample values are marked with asterisks. These plots visualize the advantage of NHPG: global maximums of the multimodal distributions is achieved close to the actual values.}
	\label{Forecasts_model_select}
\end{figure}

 In Table~\ref{Forecasting_case_variable}, we report the forecasting criteria scores. The NHPG had the best performance according to all forecasting criteria. Supplementary to Table~\ref{Forecasting_case_variable} are the plots in Figure~\ref{Forecasts_model_select}. This figure shows the approximation of the empirical posterior predictive distributions (based on a normal kernel) of the four competing models, for three randomly selected out-of-sample periods and the actual observed values in the same graph. Figure~\ref{Forecasting_case_variable} gives a graphical indication of the improved forecasting performance of NHPG.
\begin{table}[H]
	\centering
	\setlength{\tabcolsep}{20pt}
	\renewcommand{\arraystretch}{1.5}
	\begin{tabular}{lcccc}
\clrtwo 
&\multicolumn{4}{c}{\b Forecasting Criteria}\\
		
		&NHPG & M\&D &HHMM &BASAD\\
		\clrone
		CRPS&\textbf{-1.9526}&-3.6829&-2.6597&-2.4952\\
		
		MAFE&\textbf{3.9271}&4.3911&5.4101&5.0611\\
		\clrone
		MSFE&\textbf{32.8856}&39.4958&53.8280&49.1432\\  \\[-0.5cm]
		\colarray\hline
	\end{tabular}
	\caption{Forecasting performance of the competing models. In addition to M\&D and the BASAD, we include also the Homogeneous Hidden Markov Model (HHMM). The best performance (bold values) for each criterion is achieved by the proposed NHPG model.}
	\label{Forecasting_case_variable}
\end{table}
\par
\begin{figure}[H]
	\centering\includegraphics[width=\linewidth, trim=1cm 4cm 1cm 1cm]{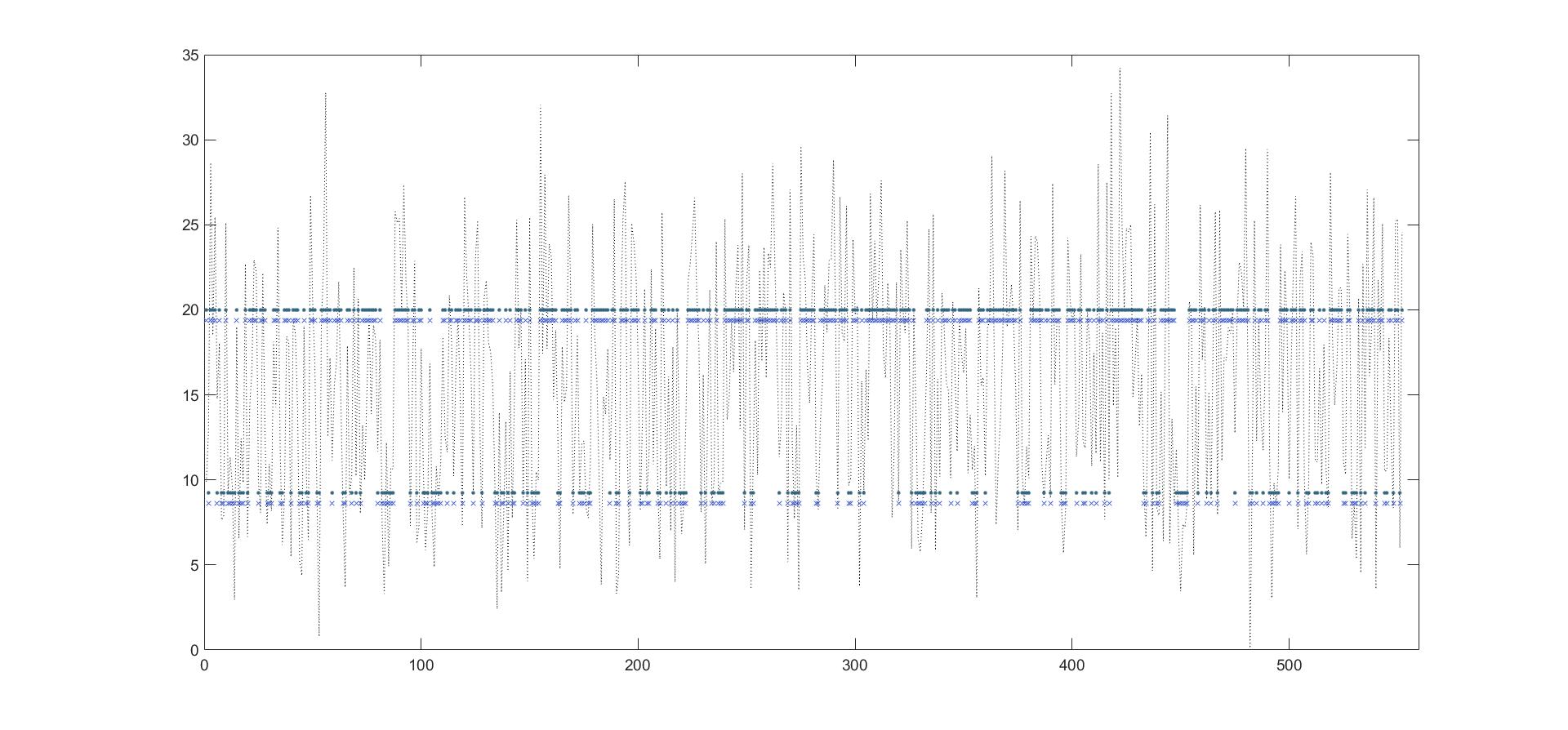}
	\caption{Observed process (black dotted line) and hidden process: the true hidden states are marked with blue x and the simulated states are marked with black dots. The true hidden process is well estimated.}
	\label{model_experiment_series}
\end{figure}

 Finally, for each MCMC iteration we kept a replicated chain of the hidden process and we compared it with the true simulated chain. Using the 0-1 Loss function, we computed the average number of misestimated states in each chain. All three approaches had similar performance according to this criterion. Specifically, from the chain with 1104 hidden states, NHPG failed to recognize 2 states per iteration, M\&D 3 states per iteration and BASAD methodology 1 state per iteration. A virtualization of the estimation of the hidden process against the true hidden process is shown in Figure~\ref{model_experiment_series}. This figure presents the thinned version (1:2 observations) of the simulated time series along with the true hidden process and an estimate of the hidden process using the proposed methodology. 
\par 
In Table~\ref{Runtime}, we report the runtimes for every methodology. The trade-off for the better forecasts of NHPG is 150 seconds per 1000 iterations in comparison to the BASAD. However, the NHPG is more than two times faster than the M\&D.

\begin{table}[H]
	\centering
	\setlength{\tabcolsep}{20pt}
\renewcommand{\arraystretch}{1.3}
	\begin{tabular}{l c c c}
	\clrtwo
		&\b NHPG &\b M\&D &\b BASAD\\
		 \specialcell[c]{ Mean runtimes per \\[-0.2cm] $1000$ iterations (seconds)} &310&693&160\\
		\colarray\hline

	\end{tabular}
	\caption{Summary of runtimes (in seconds per 1000 iterations).} 
	\label{Runtime}
\end{table}

\section{Empirical Application: Realized volatility data}\label{Empirical Application}
 We use the NHPG to assess the predictive ability of 13 financial variables in forecasting future volatility. Financial volatility has been extensively studied in the literature due to its crucial role in various financial fields, such as asset pricing, risk management, investment and asset allocation among others, see \cite{Go06}. Several studies have considered predicting realized stock volatility using various financial and/or economic predictors (see for example, \cite{Mi15, Me19, Ch12, Pa12}).
\subsection{The data}
We used the realized stock market volatility data and more precisely the \enquote{long} sample of the U.S. equity market, S$\&$P500, as described in \cite{Ch12}. The realized volatility is the squared root of the realized variance for asset class $i$ in month $t$ expressed as the sum of squared intra-period (daily) returns
$$RV_{i,t}=\sqrt{\sum_{\tau=1}^{u_t}r_{i,t,\tau}^{2}}, \; t=1\dots, T,$$ 
where $r_{i,t,\tau}$ is the $r$-th daily continuously compounded return of month $t$ for asset $i$ with $u_t$ the trading days. Thus $\sum_{\tau=1}^{u_t}r_{i,t,\tau}^{2}$ is the realized variance for asset class $i$ in month $t$. The distribution of the realized daily variances are highly non-normal and skewed to the right, but the logarithms of the realized variances are approximately normal and thus, they have better behavior (\cite{An03}). Hence, in the following analysis, we study the natural logarithm of the realized volatility series, $\ln(RV_{i,t})\; t=1\dots, T.$
\par The data are observed in a monthly basis, from December 1926 to December 2015. We used a five-years extended dataset compared to the dataset of \cite{Ch12}. The out-of-sample forecast evaluation period was set to eight years, i.e., 96 observations from December 2007 until December 2015. We had a burn in period of  $60000$ iterations and we generated $40000$ MCMC draws. We used non-informative priors for the unknown parameters $\sigma^2_s, B_s,\beta_{s}, s=1,2$, that is $\sigma _{s}^{2}\sim \mathcal{IG}(0.15,0.15)$, $B_{s}\mid \sigma _{s}^{2}\sim \mathcal{N}(0,100\sigma _{s}^{2}\times I)$ and finally $\beta_{s} \sim \mathcal{N}\(0,100\times I\)$.
\par Following \cite{Ch12} and \cite{Me19}, we took into account 13 macroeconomic and financial standardized predictive covariates. Particularly, from a list of equity market variables and risk factors, we considered the dividend price ratio (DP) and the earnings price ratio (EP) (\cite{We08}). To capture the leverage effect, i.e. the asymmetric response of volatility to positive and negative returns (\cite{Ne91}) we included the lagged equity market returns (MKT). We also used the risk factors of \cite{Fa93}, that is, the size factor (SMB), value factor (HML) and a short-term reversal factor (STR). From the set of interest rates, spreads and bond market factors, we included the treasure bill rate (TBL), i.e., the interest rate on a three-month Treasure bill, the long-term return (LTR) on long-term government bonds, the term spread (TMS), i.e., the difference between the log-term yield and treasure bill rate, the relative T-bill rate (RTB) as the difference between T-bill rate and its 12-month moving average and the relative bond rate (RBR), as the difference between LTR and its 12 month moving average (\cite{We08}). To proxy for weighted credit risk, we also used the default spread (DEF) defined as the yield spread between BAA and AAA rated bonds. Lastly, we considered the macroeconomic variable inflation rate (INF), which is the monthly growth rate of CPI. 
\par The strong contemporaneous relation between the volatility and the business conditions implies that lagged volatility plays an important role in forecasting (see \cite{Pa12,Ba19,An03}). Besides, quoting \cite{Ch12}, we include at least one autoregressive term, ``since volatility is fairly persistent, it is important to include autoregressive terms in the predictive regression to investigate whether there is additional predictive content of the macroeconomic and financial variables that goes beyond the information contained in lagged volatility''. We ran a series of experiments for this data. Specifically, we performed our analysis using the predictors described and then we repeated the analysis using the predictors plus autoregressive terms (AR) of lag 1, 2 and lag 3.
\begin{figure}[h]
	\centering
	\includegraphics[width=\linewidth, trim=0cm 4cm 0cm 0cm]{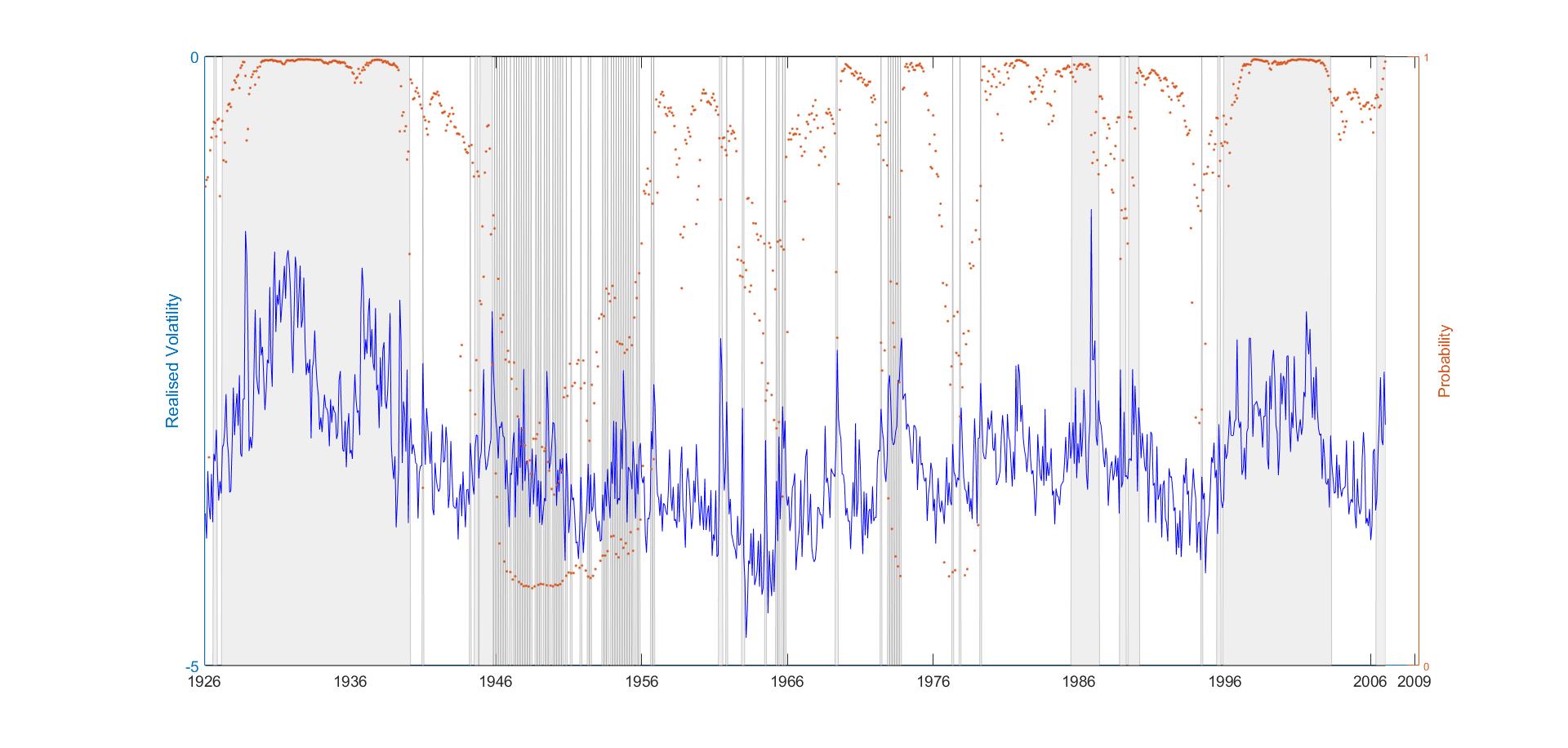}
	\caption{Time series (blue line) of the monthly realized volatility of the Standard $\&$ Poor (S\&P) 500 index (in logarithmic scale, left axis) for the period 1926-2007, using the $\text{NHPG}_1$. Gray-shaded bars mark times with hidden state 1 (smoothed probability above 0.5). The $\text{NHPG}_1$ exploits the heteroscedasticity of the series. Red dots are the posterior mean probabilities (right axis) of staying at the same state and indicate a persistent unobserved process.}
	\label{RV1}
\end{figure}
\subsection{Results}
Based on the posterior probabilities of inclusion, we see that if we do not include any Auto-Regressive (AR) terms in the predictors' pool, then the median $\text{NHPG}_0$ model has three predictors affecting only the mean equation of the series (Table~\ref{RV_Model_Selection}). Thus, based on the median probability model, the realized volatility series is considered to be a homogeneous hidden Markov model. The probabilities of staying at the same state are in this case high, concluding that the states are highly persistent. When we add the AR(1) term, the included predictors in the median probability model ($\text{NHPG}_1$) are also three but they affect the series both linearly and not linearly. We observe that an autoregressive term explains a big fraction of the variance of the realized volatility. Adding more AR terms (of lag 2 and lag 3), the median probability model remained almost the same as in the case of the model with one AR term and hence, we only report the $\text{NHPG}_1$ model. Furthermore, in our out-of-sample analysis, we did not encounter any significant improvement in the forecasting ability of the models with AR(2) and AR(3) terms. We note that this result confirms the findings of the model of \cite{Ch12}, hereafter CSS, who also used only an AR(1) term their analysis.
\par Even though -- based on the CRPS -- the model with the best performance was the one with the $\text{NHPG}_1$, we present the results of both the model with no AR terms ($\text{NHPG}_0$) and the model with one AR term ($\text{NHPG}_1$), for the sake of completeness. Also, we compare our results with those of the CSS model -- which is a linear model with one autoregressive term and a Markov Chain Monte Carlo model compositions algorithm ($\text{MC}^3$) with a Bayesian Model Averaging (BMA) approach. For the CSS model, we allowed for much longer burn in period, as suggested by the authors, of 500000 draws. Moreover we included in our comparative analysis the M\&D model and HHMM with one AR term.
\par
\begin{figure}[H]
	\centering
	\includegraphics[width=\linewidth, trim=0cm 4cm 0cm 0cm]{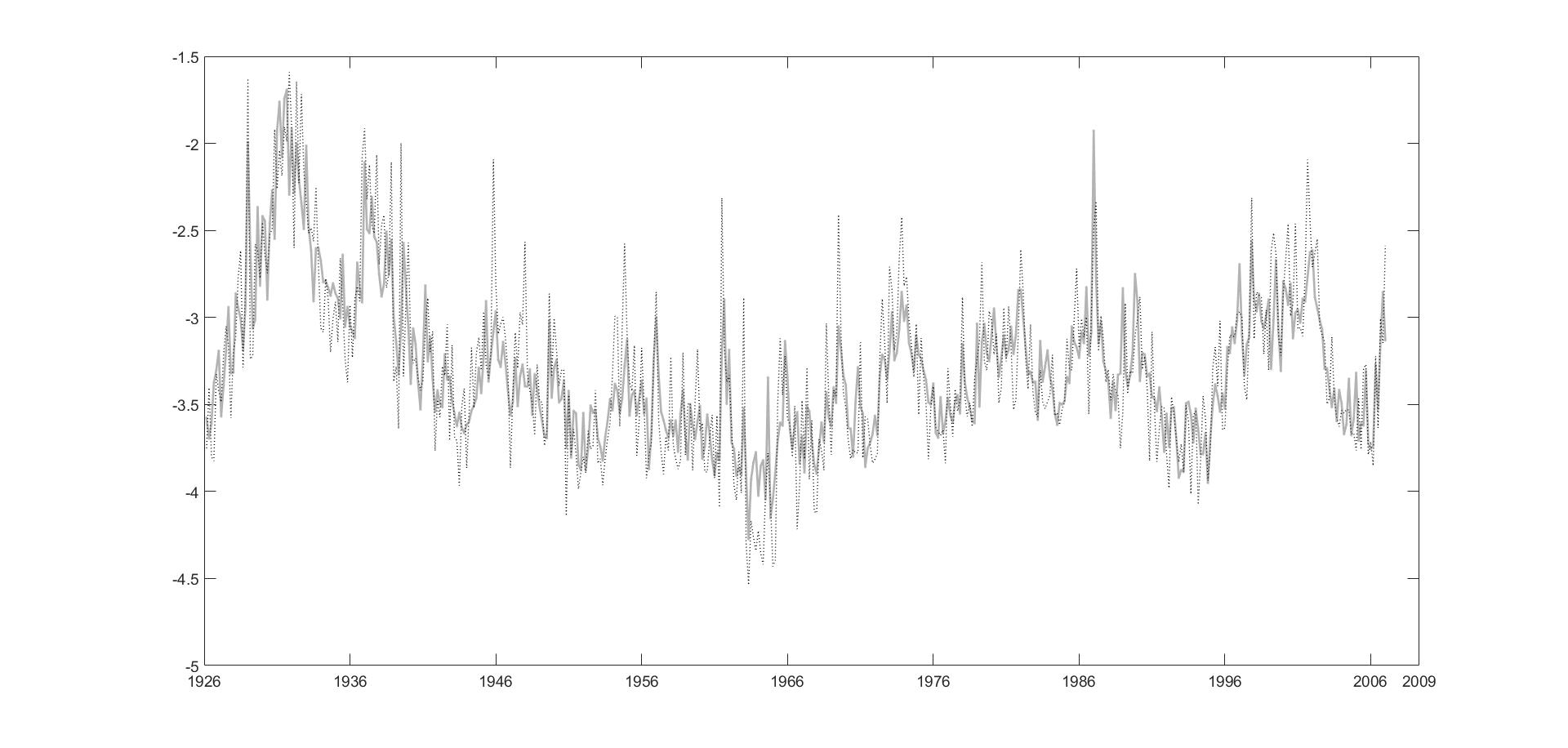}
	\caption{Thinned (1:5 observations) realized volatility time series (blue line) versus the observed process as calculated by the proposed $\text{NHPG}_1$ (gray solid line).}	\label{RV}
\end{figure}

 
Figure~\ref{RV1} shows a plot of the realized volatility data (blue line) together with the probabilities of staying at the same state (e.g. if at time $t$ we are at state 1 then the red dot at time $t$ shows the probability of staying at state $1$ at time $t+1$, that is the transition probability $p^t_{11}$). The high probabilities of staying at the same state indicate that the unobserved process is persistent. The shaded bars represent the time period that the chain was in state 1, based on the smoothed probabilities of being above 0.5 for $\text{NHPG}_1$. Furthermore, in Figure~\ref{RV} we present a thinned (1:5) in-sample realization of the observed process inferred by our algorithm, i.e., using the in-sample estimations of the parameters and the states to reproduce the realized volatility series, along with the real data. The in-sample evaluation of the observed process gives an indication of the good performance of the estimation procedure. 

\subsubsection{Model Selection}
Our model selection algorithm did not assign high probability to any specific model indicating that there exists model uncertainty. In Table~\ref{RV_Model_Selection}, we summarize the posterior probabilities of inclusion for each predictor, both for the mean equation and for the transition matrix for the $\text{NHPG}_0$, $\text{NHPG}_1$, M\&D and only for the mean equation for CSS and HHMM. Our methodology -- when the autoregressive term of lag 1 was included -- was not only able to identify which covariates affect the realized volatility series but also to decide how the covariates affect the series, i.e., linearly or non-linearly. The number of the predictors defining the median probability $\text{NHPG}_1$ has diminished to three (instead of thirteen). Specifically, we found that the MKT affects the series linearly, the SMB affects the series non-linearly and the DEF both linearly and non linearly. The predictors that were included in $\text{NHPG}_1$ are in common with the predictors included in the HHMM. The CSS model identifies four predictors with probability at least $0.5$, three of them being the same with $\text{NHPG}_1$ and HHMM, that is the MKT, the DEF and the EP plus the predictor STR. However, the M\&D algorithm includes all the predictors in the mean equation model, while it includes the predictors DP, MKT, TBL and DEF in the logistic regression for the transition probabilities. 
\begin{table}[H]
	\centering
	\setlength{\tabcolsep}{10pt}
\renewcommand{\arraystretch}{1.3}
	\begin{tabular}{lllllllll}
\clrtwo
		\multicolumn{1}{c}{}&\multicolumn{8}{c}{\b Posterior probabilities of inclusion}\\
		
		\multicolumn{1}{c}{}& \multicolumn{2}{c}{$\text{NHPG}_0$}&\multicolumn{2}{c}{$\text{NHPG}_1$} &\multicolumn{2}{c}{M\&D} &HHMM &CSS \\
		
		\clrtwo
		\b Covariates &ME&TM &ME &TM &ME & TM&ME&ME\\
		
		DP&0.01&0.04&0.08&0&\textbf{1}&\textbf{0.77}&0.3&0.38\\  
		\clrone
		EP&\textbf{0.98}&0.12&0.06&\textbf{0.89}&\textbf{1}&0.37&\textbf{0.78}&\textbf{0.50}\\
		MKT&\textbf{1}&0.04&\textbf{0.98}&0.03&\textbf{0.93}&\textbf{1}&\textbf{1}&\textbf{0.97}\\
		\clrone
		SMB&0&0.02&0&0.04&\textbf{0.90}&0.28&0&0.05\\  
		HML&0.01&0.03&0&0.02&\textbf{0.92}&0.09&0&0.06\\  
		\clrone
		STR&0&0.05&0&0.06&\textbf{0.92}&0.09&0&\textbf{0.53}\\
		TBL&0&0.02&0&0.10&\textbf{0.99}&\textbf{0.71}&0.03&0.10\\
		\clrone
		RTB&0&0.02&0&0.09&\textbf{0.99}&0.02&0&0.04\\
		LTR&0.01&0.03&0&0.03&\textbf{1}&0.01&0&0.05\\
		\clrone
		RBR&0&0.03&0&0.03&\textbf{1}&0.01&0&0.05\\		
		TMS&0.01&0.06&0.17&0.26&\textbf{0.89}&0.23&0&0.05\\
		\clrone
		DEF&\textbf{1}&0.02&\textbf{1}&\textbf{0.79}&\textbf{1}&\textbf{1}&\textbf{1}&\textbf{1}\\		
		INF&0&0.02&0.11&0.03&\textbf{0.92}&0.47&0&0.04\\
				\colarray\hline
	\end{tabular}
	\caption{Posterior probabilities of inclusion for the competing models. Predictors with inclusion probability above 0.5 (median probability model) are marked with bold values. $\text{NHPG}_0$ and $\text{NHPG}_1$ denote the proposed methodology without autoregressive terms and with one autoregressive term respectively, M\&D the methodology proposed by \cite{Me11}, HHMM the Homogeneous model with variable selection using a RJ-step and CSS the model of \cite{Ch12}. ME stands for Mean Equation (linear relationship) and TM for Transition Matrix (non linear relationship). The HHM and CSS models included covariates only in the ME. }
	\label{RV_Model_Selection}
\end{table}
\subsubsection{Forecasting}\label{Forecasting_RV} 
The values of the forecasting criteria that we used for all competing models are reported in Table~\ref{Forecasting criteria_RV}. We conclude that $\text{NHPG}_1$ performs better than all the other models, since it has the best scores in all forecasting criteria: the mean Continuous Ranked Probability Score (E(CPRS)), MAFE and MSFE. 
\begin{table}[H]
	\centering
	\setlength{\tabcolsep}{10pt}
\renewcommand{\arraystretch}{1.3}
	\begin{tabular}{lccccc}
	\clrtwo
		&\multicolumn{5}{c}{\b Forecasting Criteria}\\
		
		&$\text{NHPG}_1$ &$\text{NHPG}_0$& M\&D&HHMM &CSS\\
		\clrone
		CRPS&\textbf{-0.1971}&-0.2175&-0.2191&-0.2118&-0.2238\\
		
		MAFE&\textbf{0.3821}&0.4643&0.4172&0.4534&0.4787\\
		\clrone
		MSFE&\textbf{0.2467}&0.3426&0.2678&0.3449&0.3813\\ \\[-0.5cm]
		\colarray\hline
	\end{tabular}
	\caption{Summary of forecasting results of the five competing models, obtained from the log-realized volatility dataset. The best performance (bold values) for each criterion is achieved by the $\text{NHPG}_1$ model (with one autoregressive term).}
	\label{Forecasting criteria_RV}
\end{table}
\section{Conclusions}\label{Conclusions}
In this paper, we considered inference on predictive Non-Homogeneous Hidden Markov Models with P\'{o}lya-Gamma data augmentation. Given a common pool of predictors, we allowed for different sets of covariates to affect the mean equation and the time-varying transition probabilities. To determine which covariates affect the series linearly and/or non-linearly, we performed stochastic variable selection using a double reversible jump step. Additionally, we modeled the probabilities of the transition probability matrix via a logistic link. Bayesian inference for the logistic regression model has been recognized as a hard problem -- many of the proposed methodologies face efficiency and convergence issues -- due to the analytically inconvenient form of the model's likelihood function. To account for these issues, which are amplified in the more complex setting of NHHMMs, we developed an accurate MCMC inference scheme in this setting, using the recently proposed P\'{o}lya-Gamma data augmentation scheme of \cite{Po13}. 
\par In each MCMC iteration, we simulated the hidden states using the scaled Forward-Backward algorithm of \cite{Sc02}, the mean equation parameters using a Gibbs step, and the logistic regression coefficients using the P\'{o}lya-Gamma augmentation scheme. Finally, we performed a double reversible jump step to choose the covariates that affect the mean equation and the transition probabilities. Using the most probable model, the median probability model or Bayesian Model Averaging, we make one-step-look ahead predictions, within the Bayesian framework. \par
To assess the performance of the proposed algorithm and the predictive ability of our model, we conducted an extensive number of simulation experiments. The results showed that our algorithm mixes and converges well and provides accurate estimates of the model's parameters. Moreover, they exhibited that our model outperforms benchmark models, such as the approach of \cite{Me11}, the BASAD model of \cite{Na14} and the homogeneous hidden Markov model, in terms of both variable selection and forecasting ability according to the continuous ranked probability score, the mean absolute forecasting error and the mean square forecasting error. The currently proposed methodology was applied to a realized volatility dataset -- detailed in \cite{Ch12} -- for predicting future observations and for predictor selection. The median probability model identified three predictors, one affecting the analyzed series linearly, one non-linearly and one both linearly and non-linearly. Using the proposed methodology we obtained improved forecasts, compared to \cite{Ch12}.
\par
The findings of the present study indicate that complex Non-Homogeneous Hidden Markov models are promising for predicting univariate financial and economic time series. More accurate forecasts can be derived without the need of tuning (black box functionality) and at a low trade-off in terms of computational complexity. The efficiency of the proposed model can be further improved by refining the model selection process. Moreover, using standard methods, it can be extended to the prediction of multivariate time series that arise in many economic and non-economic applications. In this way, the proposed methodology may be of interest not only to the econometric but also to the broader forecasting community.

\section*{Data Availability Statement}
The data that support the findings of this study are available from Journal of the Operational Research Society. Restrictions apply to the availability of these data, which were used under license for this study. Data are available from Ekaterini Panopoulou/ at https://doi.org/10.1080/01605682.2018.1489354 with the permission of Journal of the Operational Research Society.

\section*{Appendix}\label{Appendix}
\subsection{Benchmark models}\label{ApBench}
We give the definitions of the BASAD model, \cite{Na14}, and the standard Homogeneous Hidden Markov Model (HHMM) that we use in Section~\ref{Variable selection}. In the BASAD model, the authors introduce shrinking and diffusing priors as a spike and slab priors model, with prior parameters depending on the sample size to achieve appropriate shrinkage. They work with orthogonal design matrices and use binary latent variables $U_i$ to indicate if a covariate is active or not. In our setting, the BASAD model for the mean equation is defined as:
$$Y_t\mid \(X_{t-1} B_s,\sigma^2_s\) \sim \mathcal{N}\(X_{t-1}B_s,\sigma^2_{s}\),\ s=1,2, \; t=1,\dots , T,$$
$$B_{k,s}\mid \(\sigma^2_s,U_{k,s}=0\) \sim \mathcal{N}\(0,\sigma^2_{s}\tau^{2}_{0B,n}\),\ B_{k,s}\mid \(\sigma^2_s,U_{k,s}=1\) \sim \mathcal{N}\(0,\sigma^2_{s}\tau^{2}_{1B,n}\),\  k=1,\dots,r,$$
$$P\(U_{k,s}=1\)=1-P\(U_{k,s}=0\)=q_n, \  k=1,\dots,r,$$
$$\text{and}$$
$$\sigma^2_s \sim \mathcal{IG}\(\alpha_1,\alpha_2\).$$
The transition probabilities are parametrized as:
$$\tilde{Z}^{s}_{t+1} \sim Bin\(1,\frac{\exp\(x_{t}\beta_{s}\)}{1+\exp\(x_{t}\beta_{s}\)}\),$$
$$\beta_{k}^{s}\mid \(U_{k,s}=0\) \sim \mathcal{N}\(0,\sigma^2_{s}\tau^{2}_{0\beta,n}\), \ \beta_{i}^{s}\mid\(U_{k,s}=1\) \sim \mathcal{N}\(0,\sigma^2_{s}\tau^{2}_{1\beta,n}\), \ k=1,\dots,r,$$
$$\omega_{s} \sim \mathcal{PG}\(b_{\omega},0\)$$
$$\text{and}$$
$$P\(U_{k,s}=1\)=1-P\(U_{k,s}=0\)=q_n, \ k=1,\dots,r.$$
In contrast, in the HHMM, covariates affect only the mean equation and the transition probability matrix is constant, 
$$Y_t\mid Z_t=s \sim  \mathcal{N}\(X^{(1)}_{t-1}B_s,\sigma^2_{s}\), \ s=1,2, \; t=1,\dots , T,$$
$$P\(z_t=j\mid z_{t-1}=i\) =p_{ij}, \; i,j=1,2 \; \forall \; t=1,\dots T.$$	

\subsection{Case study for the fixed model}\label{ApCaseStudy}
We present the results of a case study without model uncertainty. This study shows empirically that our algorithm converges, mixes well and is effective. In this case, the results are marginally better than the M\&D model and the HHMM. However, together with the results of the case with model uncertainty, they demonstrate that the proposed algorithm provides an overall improvement over M\&D.
\par
We simulated data from a NHHMM of size $T=1500$. We used three covariates $X^{(1)}=\left\{1,X_1,X_2,X_3\right\}$ affecting the mean equation and three covariates $X^{(2)}=\left\{1,X_1,X_2,X_4\right\}$ affecting the transition matrix, with $X$ independently normally distributed covariates with means $\mu_x=\left[4,3,-2,-5\right]$ and variances $\sigma^2_x=\left[1,1,0.5,1\right]$. The mean equation parameters were $B_1=\left[2, -0.3,2, 2\right]',\ \sigma_1^2=1.5$ and $B_2=\left[1, 3, 4, 3\right]',\ \sigma_2^2=0.8$ whereas the logistic regression coefficients where $\beta_1=\left[1.5,1,2, 3\right]'$ and $\beta_2=\left[3, -2.5, 4, 1\right]'$ for the two states, respectively. We kept $L=100$ out-of-sample observations and we computed a sequence of one-step-ahead forecasts of the real observed process. We used non-informative priors for the unknown parameters $\sigma_s, B_s,\beta_{s},\ s=1,2$, that is $\sigma _{s}^{2}\sim \mathcal{IG}(0.1,0.1)$, $B_{s}\mid \sigma _{s}^{2}\sim \mathcal{N}\(0,100\sigma^2_s\times I\)$ and finally $\beta_{s}\sim \mathcal{N}\(0,100\times I\)$.
\begin{table}[H]
	\centering
	\setlength{\tabcolsep}{10pt}
	\renewcommand{\arraystretch}{1.3}
	\begin{tabular}{llrrr}
		\clrtwo
		&&\b NHPG &\b M\&D &\b HHMM \\
		
		Forecasting criteria &CRPS&\textbf{-2.0794}&-2.1920&-4.4323\\
		
		&MAFE&\textbf{4.2661}&4.3907&8.6745\\
		
		&MSFE&\textbf{60.9978}&63.1617&143.1512\\  
		\clrone
		Sample Quality&ESS & 11936&11934&\textbf{12153}\\
		\clrone
		&mESS & 24391& 24343&\textbf{24578} \\
			
		Efficiency&mESR& 3.2130& 0.0120&\textbf{110}\\
		\clrone
		Convergence \& Mixing&PSRF&\textbf{1}&\textbf{1}&\textbf{1.0005}\\
		\clrone
		&mCM &\textbf{18}(18)&\textbf{18}(18)&8(10)\\
		\colarray\hline
		
	\end{tabular}
	\caption{Summary of results: CRPS is the mean continuous rank probability score, MAFE is the mean absolute forecast error and MSFE is the mean square forecast error. ESS is the minimum effective size of among the ESS for all parameters and mESS the multivariate effective size, for an MCMC run of 25000 iterations. mESR is the minimum effective sample rate. PSRF is the maximum potential scale reduction factor and mCM is the multivariate convergence and mixing diagnostic. In the mCM line we report the number of the components of the parameters out of the total components -- in parenthesis -- that fall into the 95\% confidence interval of the test. NHPG is the proposed model, M\&D is the model of \cite{Me11}, HHMM is the homogeneous model. Bold values denote the best values for the corresponding criterion among all the competing models.}
	\label{Summary_fixed}
\end{table}
\par
\begin{figure}[H]
	\centering
	\includegraphics[width=\linewidth, trim=0cm 4cm 0cm 0cm]{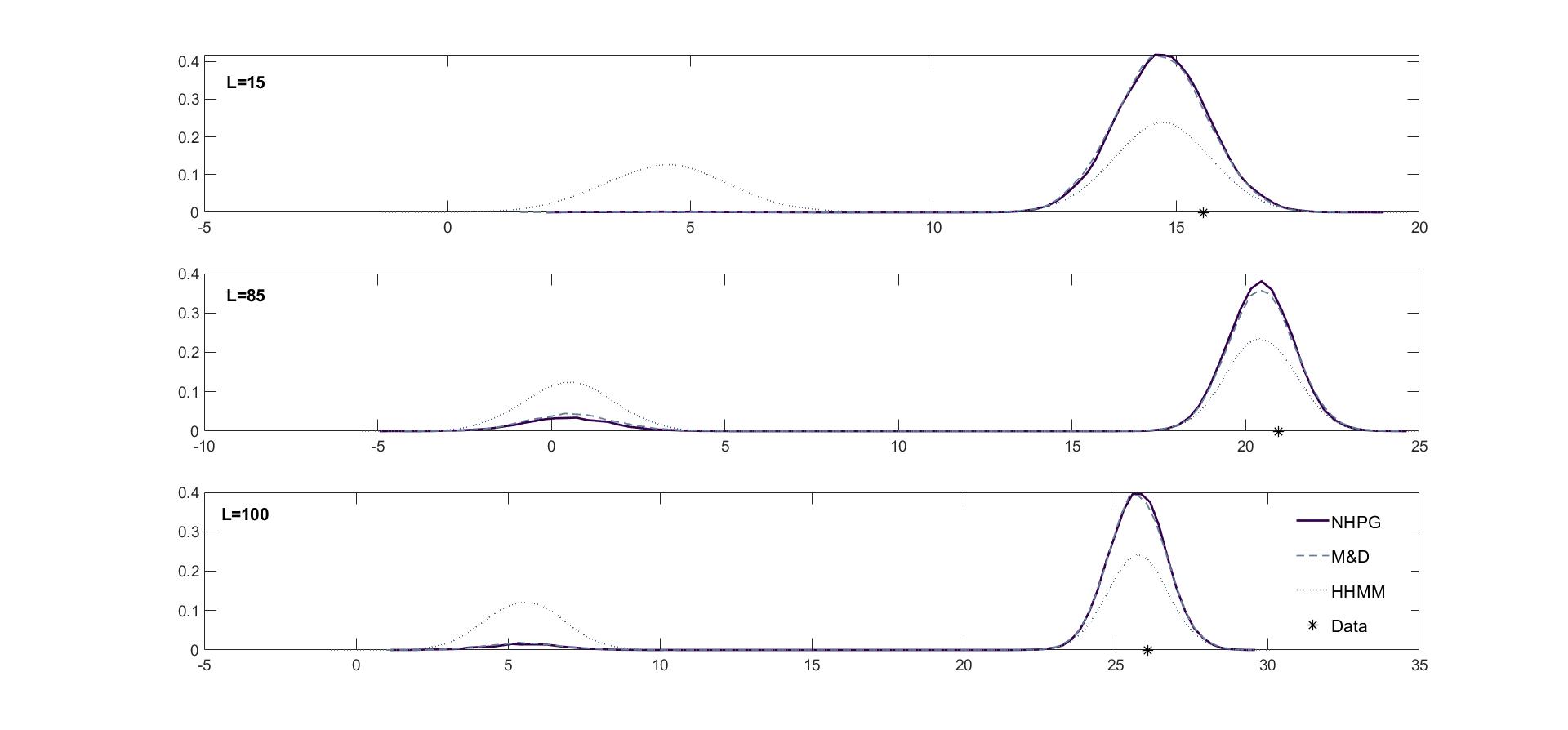}
	\caption{Conjointly plotted empirical continuous approximations (based on a normal kernel function) of the posterior predictive distribution for out-of-sample periods $L=15,85,100$, using the NHPG (black continuous line), M\&D (gray dashed line) and the HHMM (gray dotted line). Actual out-of-sample values are marked with asterisks.}
	\label{Exp1F}
\end{figure}

 Inferences are based on an MCMC sample of 25000 iterations after a burn-in period of 10000 iterations. A summary of the results of this experiment is reported in Table~\ref{Summary_fixed}. We used several metrics for assessing the efficiency of our algorithm (see Appendix~\ref{ApCriteria}). The quality of the sample is measured with the effective sample size (ESS), multivariate effective sample size (mESS). We also use the minimum Effective Sample Rate (mESR) as a measure of the efficiency of the algorithm. To assess the convergence and mixing of the algorithm we use the Potential Scale Reduction factor (PSRF) of \cite{Br98,Ge13} and the multivariate mixing diagnostic of \cite{Pa12}. Specifically, we show that our algorithm converges to the stationary distribution and has good mixing properties, using the aforementioned diagnostic criteria. Also, the univariate and multivariate effective sample sizes for all the methodologies/models are large, implying an efficient algorithm. The Effective Sample Rate of NHPG is $3.213$ whilst M\&D's ESR is significantly lower with a score of $0.012$. In addition, the NHPG has best forecasting performance, since it has the best score among the benchmarks' scores in all forecasting criteria. In Figure~\ref{Exp1F} we visualize the empirical continuous approximation of the posterior predictive densities of NHPG, M\&D, HHMM, for the three randomly selected out-of-sample periods, $L=15,85,100$. These plots provide additional evidence that the NHPG gives at least good predictions as the M\&D model. 
\subsection{The Continuous Rank Probability Score}\label{ApCRPS}
Let $y_{l}$ be the real observed values of the forecasts, $\textbf{y}$ the history of the predictive quantity and $\hat{y}_{l}$ the estimated forecasts. Using the notation  $f$ for the distribution of the true model, $f_p(\hat{y}_l)$ for the  posterior predictive density of the new data, and $F_p(x)=\int_{-\infty}^{x}f_p\(\hat{y}\mid \textbf{y}\)d\hat{y}$  for the posterior predictive cumulative density function. The CRPS for $y_i$ is defined as,
$$CRPS(F_{p,l},y_l)=-\int_{-\infty}^{\infty}\(F_{p,l}\(\hat{y}_l\)-F_{y_i}\(\hat{y}_l\)\)^2 d\hat{y}_l,$$
where $I\(x\geq y\)$ denotes a step function along the real line that attains the value 1 if $x\geq y$ and the value 0 otherwise, $F_{y_l}=H(\hat{y}_l-y_{l})$ is the cumulative distribution of the real value $y_l$ and $H$ is the Heaviside function (\cite{He00}), $H(x) =0$, if $x\leq 0$ and $1$ otherwise. 
\subsection{Metrics of Comparison}\label{ApCriteria}
\par We briefly present the convergence diagnostics, mixing criteria and metrics of effectiveness that we used to measure the performance of our algorithm.
\par 
As a primary metric of comparison, following \cite{Ho06} and \cite{Po13}, we calculated the effective sample size (ESS). For each dimension of the parameter vector, $\text{ESS}_i$ is the number of independent samples needed to obtain a parameter estimate with the same standard error as the MCMC estimate based on $M$ dependent samples (see \cite{Ne93,Ka98}). If $\theta$ is the $p-$dimensional parameter of interest, and $\theta_n=1/n\sum_{t=1}^{n}g(x_t)$ is an estimate of $\theta$ based on a Markov chain $\{X_t\}$, with $\theta_n\longrightarrow \theta$ the Monte Carlo error, $\theta_n - \theta$ is described asymptotically by the Central Limit Theorem (CLT), $\sqrt{n}\(\theta_n-\theta\)\xrightarrow[n\rightarrow \infty]{d}\mathcal{N}\(0,\Sigma_p\)$. The idea of the EES lies on the univariate CLT for each component of $\theta$ and it is defined, for $i=1,\dots,p$, as 
$$\text{ESS}_i=\frac{M}{1+2\sum_{j=1}^{k}\rho\(j\)}=M\frac{\lambda_i}{\sigma_i},$$ 
where $\rho\(k\)$ is the sample autocorrelation of lag $k$ of the parameter $\theta_i$, $\lambda_i$ the diagonal element of the sample covariance matrix $\Lambda$, $\sigma_i$ the diagonal element of $\Sigma_p$ and $M$ the number of post-burn in samples. We also report the minimum Effective Sample Rate (mESR) to compare a slow sampler with a fast sampler, as in \cite{Po13,Fr10}. The mESR is defined as the minimum ESS per second of running time, i.e., $\text{mESR}=\(\text{mESS}\)/t_{\text{cpu}}$. It quantifies how rapidly a Markov-chain sampler can produce independent draws from the posterior distribution.
\par \cite{Va19} argue that a univariate approach ignores cross-correlation across components, leading to an inaccurate picture of the quality of the sample. Thus, they define a multivariate version of the ESS. Specifically, $$\text{mESS}=M\(\frac{\arrowvert\Lambda\arrowvert}{\arrowvert\Sigma_p\arrowvert}\)^{1/p}.$$
When there is no correlation, then $\Sigma=\Lambda$ and $\text{mESS}=M$. 
\par To assess the convergence and mixing of our algorithm we use the Potential Scale Reduction factor (PSRF) of \cite{Br98,Ge13} and the multivariate convergence and mixing diagnostic proposed by \cite{Pa12B}. In brief, implementation of PSRF requires sample runs from multiple chains (alternatively a very long chain can be divided into two or more subchains). The key quantity is the ratio of the resulting between- and within-chain variances. If the within-chain variance dominates the between-chain variance, the ratio approaches 1, which suggests that the chains have approximately reached stationarity. Desirable values for PSRF are the values below $1.1$ for every component of the parameters. In short, \cite{Pa12B} obtain MCMC-based estimators of posterior expectations by combining different subgroup (subchain) estimators using stratification and post-stratification methods. They develop variance estimates of the limiting distributions of these estimators. Based on these variance estimates, they propose a statistic test to aid in the assessment of convergence and mixing of chains.
\bibliographystyle{abbrvnat}
\bibliography{mybib}

\end{document}